\documentclass[12pt]{article}
\usepackage{amsmath,amssymb,amsthm,mathrsfs,bm}
\usepackage{amscd}
\usepackage{wrapfig}
\usepackage{tcolorbox}
\usepackage{float}
\usepackage{makeidx}
\usepackage[all]{xy}
\usepackage{url}
\usepackage[dvipdfmx]{hyperref}
\theoremstyle{definition}

\newtheorem*{thm*}{Theorem}

\newtheorem*{defn*}{Definition}

\newtheorem*{lem*}{Lemma}

\newtheorem*{rem*}{Remark}

\newtheorem*{con*}{Conjecture}

\newtheorem*{cor*}{Corollary}

\newtheorem*{prop*}{Proposition}

\newtheorem*{hypoth*}{Hypothesis}

\newtheorem*{claim*}{Claim}

\makeindex

\newcommand*{\affaddr}[1]{#1} 

\newcommand*{\email}[1]{\texttt{#1}}
\begin{document}
\title{\vspace{2cm}\bf\Large{\textsf {Quantum Hall Effect and Langlands Program}\\
\noindent\rule{\textwidth}{1.5pt}}}
\author{%
\bf{\textsf Kazuki Ikeda}\\
\affaddr{\small{Department of Physics, Osaka University, Toyonaka, Osaka 560-0043, Japan}\\\small{\email{kikeda@het.phys.sci.osaka-u.ac.jp}}}
}
\date{}
\maketitle
\begin{abstract}
\hspace{-6mm}Recent advances in the Langlands program shed light on a vast area of modern mathematics from an unconventional viewpoint, including number theory, gauge theory, representation, knot theory and etc. By applying to physics, these novel perspectives endow with a unified account of the (integer/ fractional) quantum Hall effect. The plateaus of the Hall conductance are described by Hecke eigensheaves of the geometric Langlands correspondence. Especially, the particle-vortex duality, which is explained by $S$-duality of Chern-Simons theory, corresponds to the Langlands duality in Wilson and Hecke operators.  Moreover the Langlands duality in the quantum group associated with the Hamiltonian describes fractal energy spectrum structure, know as Hofstadter's butterfly. These results suggest that the Langlands program has many physically realistic meanings. 
\end{abstract}
\newpage
\section{Introduction} 
Ramanujan's finding on automorphic forms is crucial to modern number theory. Around 1916, he calculated the expansion coefficients $a_n$ of the following infinite series. 
\begin{equation}\label{1954}
q\prod_{n=1}^\infty (1-q^n)^2(1-q^{11n})^2=\sum_{n=1}a_nq^n.
\end{equation}
About 40 years later, Eichker proved that there is a profound correspondence between the automorphic form above and the elliptic curve defined on $\mathbb{Q}$
\begin{equation}\label{eli}
y^2+y=x^3-x. 
\end{equation}
Astonishingly, for $b_p=p+1-\#~(\text{points of}~\eqref{eli}~\text{mod}~p)$, the equation
\begin{equation}
a_p=b_p
\end{equation}
is true for any prime $p$. It is the Langlands program \cite{Langlands1970} that connects these dualities from general viewpoints of mathematics. In the Langlands program, the correspondence between automorphic forms and elliptic curves are all about the correspondence between the eigenvalues of Hecke operators and Frobenius operators, namely $a_p$ and $b_p$ are eigenvalues of Hecke and Frobenius operators respectively. The Langlands program can be interpreted geometrically \cite{Deligne1973,Drinfel'd1987,Lafforgue2002}, and the geometric Langlands correspondence foresees many nontrivial aspects of gauge theories. From a perspective of geometry, in a simple case, this is achieved by considering a gauge theory on a Riemann surface, and a Hecke operator modifies the given principle bundle at a singular point so that the 1st Chern-number jumps at the singularity, as a vortex or a monopole operator do, and a Frobenius operator is analogous to a holomony operator, like a Wilson loop \cite{Frenkel:2005pa,Kapustin:2006pk,Witten:2015dta}. 

The aim of this article is to endow it with physical meaning. In a seminal piece of research made by A. Kapustin and E. Witten \cite{Kapustin:2006pk}, they predict the electric magnetic duality and mirror symmetry are intimately related to the geometric Langlands correspondence. While there are many relevant works \cite{Witten:2015dta,Gaiotto:2016hvd,Gukov:2006jk,Witten:2009at}, what would follow in view of non supersymmetric physics had not been known. In this work, we address the quantum Hall effect and enjoy the panoptic picture of the quantum Hall effect drawn as a natural consequence of the Langlands program. One can seek a cue from the Langlands/GNO dual group to understand the connection with the quantum Hall effect and the geometric Langlands duality. In electric-magnetic duality, Dirac monopoles and Dirac's quantization condition explain the dual group $^LG$ of a general Lie group $G$ from a perspective of physics \cite{GODDARD19771}. The quantization condition of the Hall conductance is understand in a similar manner. 

\begin{figure}[h!]\label{hall}
\centering
\includegraphics[width =6cm]{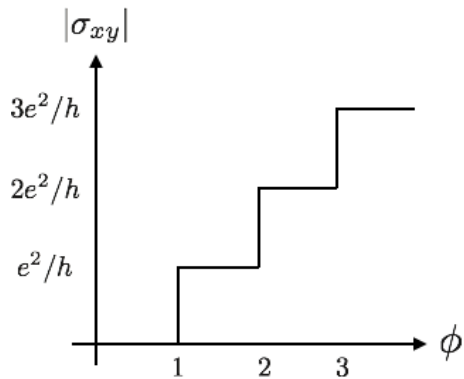}
\caption{Integer dependency of the Hall conductance}
\end{figure}

More concretely, what we address in this article is 
\begin{enumerate}
\item Integer quantum Hall effect (IQHE)
\item Fractional quantum Hall effect (FQHE)
\item Hofstadter's butterfly
\end{enumerate}
First of all, what is common in these three is that they are described by the $S$-duality like picture, as often discussed in gauge theory. They may look different at first sight, however, the Langlands philosophy connects them all eventually. The two dimensional IQHE shows the typical Hall conductance $\sigma_{xy}$ classified by integers (figure \ref{hall}).
$\sigma_{xy}$ is given by the sum of the Chern numbers\footnote{In this article, we will often call the 1st Chern number of a $U(1)$-bundle as its Chern number since we focus on two-dimensional physics. When we say Chern numbers, it means we consider many $U(1)$-bundles.} of the $U(1)$-bundles, associated with the energy bands below the Fermi level, on the Brillouin zone (BZ) \cite{PhysRevLett.49.405,kohmoto1985topological}. The plateau regions clearly explains the existence of flat connections forming $\mathcal{D}$-modules, which turn out to be Hecke eigensheves. And quantum jumps in the conductivity are naturally described by the Hecke modification. The Hall conductance $\sigma$ which accompanies the Chern-Simons action and $q$-parameter accommodates $\sigma$ in such a way that $q=\exp(\sigma)$. The duality of Chern-Simons theory induces flipping $q\to ^Lq=\exp(-1/\sigma)$ which suggests the IQHE-FQHE duality. This is how the Langlands/S-duality unites the IQHE and the FQHE. 

Secondly, it is important to recall that the Hamiltonian of the IQHE is written by the quantum group $\mathcal{U}_q(sl_2)$ and $S$-duality maps it to the dual quantum group with the dual $^Lq$-parameter. Flipping $q\to ^Lq$ corresponds to the particle-vortex duality. And we explain that Hofstadter's butterfly is captured in this way. 

A better understanding on the IQHE will also enhance comprehension of general topological insulators (TI). Roughly speaking, TI's are extension of the IQHE to general gauge theories and they are classified in a similar manner \cite{PhysRevB.78.195125}. And analogous idea of the particle-vortex duality can be seen in general ways, including supper conductors \cite{Murugan:2016zal,PhysRevB.93.245151}. While the QHE and TI's are well known systematically, geometric Langlands correspondence for general cases remains a conjecture. Hence knowledge on the QHE and TI's will endow the Langlands program with hints for being developed. 

This piece is orchestrated as follows. In section 2, we address the IQHE and the FQHE from a perspective of the geometric Langlands correspondence. In section 3, we describe the Langlands duality in the quantum group $\mathcal{U}_q(sl_2)$ and build a connection with the geometric Langladns duality based on Chern-Simons theory. Hofstadter's butterfly appears on this way. 
\paragraph*{Acknowledgement}$\\$
I thank Kentaro Nomura for useful conversations on the quantum Hall effect and topological insulators. I am grateful to Sergei Gukov for various comments. 

\section{Integer Quantum Hall Effect and Langlands Program}
\subsection{General Setup} 
Throughout this article we consider the IQHE on Laughlin's type of geometry \cite{PhysRevB.23.5632}, namely a square lattice in  the uniform magnetic flux $\phi$ perpendicular to the system which has a period $L_y\in\mathbb{Z}$ in the $y$ direction. We assume $\phi$ is rational $P/Q$, where $P$ and $Q$ are mutually prime integers, then the Hall conductance $\sigma_{xy}$ is quantized and there are $Q$ energy bands. To consider energy bands, we prefer to work on a generic tight-binding Hamiltonian 
\begin{equation}\label{H}
H=\sum_{m,n}\left(c_{m+1,n}^\dagger c_{m,n}e^{i\theta^x_{m,n}}+c_{m,n+1}^\dagger c_{m,n}e^{i\theta^y_{m,n}}+h.c.\right),
\end{equation} 
where $c_{m,n}~(c_{m,n}^\dagger)$ is the annihilation (creation) operator at $(m,n)$ site. If we choose the Landau gauge $(\theta^x_{m,n},\theta^y_{m,n})=(0,2\pi m\phi)$, the Shcr\"{o}dinger equation becomes 
\begin{equation}\label{s}
\Psi_{m+1,n}+\Psi_{m-1,n}+e^{i2\pi m\phi}\Psi_{m,n+1}+e^{-i2\pi m\phi}\Psi_{m,n-1}=E\Psi_{m,n}. 
\end{equation}
We write $\Psi_{m,n}=e^{ik_yn}\psi_{m}(k_y)~(0\le k_y\le 2\pi)$ since the system is periodic in the $y$ direction. Under the assumption that $L_y$ is sufficiently large, the wave number $k_y=2\pi l/L_y~(l\in\mathbb{Z})$ is usually regarded as a continuous parameter. Moreover the system has a period $Q$ in the $x$ direction because of the rational flux $\phi=P/Q$, therefore Bloch's theorem allows us to write the wave function as $\psi_{m}(k_y)=e^{i2\pi mk_x}u_{m}(k_x, k_y)~(0\le k_x\le 2\pi/Q)$, where $u_m$ is periodic $u_{m+Q}=u_m$. In view of the wavenumber space or the Brillouin zone, the system has two periods $2\pi/Q$ and $2\pi$ in the $k_x$ and $k_y$ directions respectively, hence we identify the BZ with a torus $T_{BZ}^2$ by gluing its boundary.  There are $Q$-energy bands\footnote{One should be careful not to confuse the band and site indexes. Their total numbers are the same.} and each of them is a $U(1)$-bundle on $T^2_{BZ}$.  A $U(1)$-connection for the $j$-th bundle is given by the Berry connection $A^j=-i\sum_{m=1}^{Q}({u_m^j}^\dagger \partial_{k_x}u_m^jdk_x+{u_m^j}^\dagger \partial_{k_y}u_m^jdk_y)$, where $u_m^j$ is the Bloch function for the $j$-th energy band and normalized $|u^j|^2=\sum_{m=1}^Q{u_m^j}^\dagger u_m^j=1$.  

\subsection{Hall Conductance}\label{H1}
One of the reasons for the quantized Hall conductance $\sigma_{xy}$ can described by the Chern numbers of the fiber bundles. Let $\sigma_{xy}^j$ be the Hall conductance of the $j$-th energy band. The well-known formula 
\begin{equation}
\sigma^j_{xy}=\frac{e^2}{h}\int_{T^2_{BZ}}\frac{d^2k}{2\pi}\left(\frac{\partial A^j_y}{\partial k_x}-\frac{\partial A^j_x}{\partial k_y}\right)
\end{equation} 
tells that the Hall conductance is given by the Chern number of the $j$-th $U(1)$-bundle $\mathcal{L}^j$. If the Berry connection $a^j$ is holomorphic on entire $T^2_{BZ}$, then the Stokes theorem implies $\sigma^j_{xy}=0$. So for $\sigma^j_{xy}$ being nontrivial, $a^j$ must have a singular point on $T^2_{BZ}$. At such a point, the Bloch function $u^j(k)$ vanishes and the Chern number corresponds to vorticity of the function \cite{kohmoto1985topological, PhysRevLett.71.3697}. If there exist many singular points, $\sigma^j_{xy}$ is given by the total vorticity. This viewpoint is important for the Langlands correspondence, especially for the Hecke modifications of bundles.  

Let us generalize the statement above. Suppose the Fermi energy $E_F$ lies in the $M$-th gap, and we write $M$ states of bands below $E_F$ by $u=(u^1,\cdots, u^M)$. This multiplet forms a $U(M)$-bundle $\mathcal{E}$ over $T^2_{BZ}$, whose Berry connection is given by $A=-i\text{Tr}(u^\dagger du)$. Then the Hall conductance $\sigma_{xy}$ is given by the Chern number $c(\mathcal{E})=\frac{1}{2\pi}\int_{T^2_{BZ}}\text{Tr}(dA)$. The famous TKNN formula \cite{PhysRevLett.49.405} tells that the total Hall conductance $\sigma_{xy}$ is given by the sum of Chern numbers $c(\mathcal{L}_j)=\frac{1}{2\pi }\int_{T^2_{BZ}} dA^j$ associated with all energy bands below $E_F$: 
\begin{equation}
\sigma_{xy}=\frac{e^2}{h}\sum_{j=1}^Mc(\mathcal{L}_j).
\end{equation}
 
\subsection{\label{GL}Geometric Langlands Correspondence}
The geometric Langlands correspondence is a branch of the Langlands program. There are a lot of surveys and the conjecture is partly proven \cite{Deligne1973,Drinfel'd1987,Lafforgue2002}.  The $GL_1=\mathbb{C}^\times$ case\footnote{The compactification of $GL_1=\mathbb{C}^\times=\mathbb{C}\setminus\{0\}$ is $U(1)$.} is the simplest and established. We will focus on this case for a while. A readable introduction is \cite{Frenkel:2005pa} whose part II will help us greatly. Let $X$ be a compact Riemann surface. We consider a holonomy representation $\rho:\pi_1(X)\to GL_1$. A famous mathematical theorem\footnote{This is true for any smooth manifold $M$ and any representation $\rho:\pi_1(M)\to G$, where $G$ is an arbitrary Lie group.} guaranties a bijection between the set $\text{Loc}_1(X)$ of isomorphism classes of flat $GL_1$-bundles on $X$ and the set of conjugacy classes of the holomony representations. Hence, one can attach a flat connection for a given representation $\rho$. (This is the same trick we define a Wilson loop.) An element of $\text{Loc}_1(X)$ is called a local system. 

Now we introduce another character of the geometric Langlands correspondence. We denote by $\text{Pic}(X)$ the set of isomorphism classes of holomorphic line bundles on $X$, which classifies the line bundles by their 1st Chern classes:
\begin{equation}
\text{Pic}(X)=\bigsqcup_{d=0}\left\{\mathcal{L}\in\text{Pic}(X):d=\int_Xc_1(\mathcal{L})\right\}.
\end{equation} 
For a given $x\in X$, we consider the map, called a Hecke operator (functor), 
\begin{align}\label{hecke}
\begin{aligned}
h_x:\text{Pic}(X)&\to \text{Pic}(X)\\
&\mathcal{L}\mapsto\mathcal{L}(x),
\end{aligned} 
\end{align}
where $\mathcal{L}(x)$ is the line bundle whose sections are sections of $\mathcal{L}$ which may vanish at $x$. Under $h_x$, the Chern number of $\mathcal{L}$ jumps by 1 $(c(\mathcal{L}(x))=c(\mathcal{L})+1)$. One can consider a more general modification of $\mathcal{L}$ to $\mathcal{L}'$ at $N$-tuple of points $(x_i),i=1,\cdots, N$ so that $c(\mathcal{L}')=c(\mathcal{L})+N$. 

What the geometric Langlands correspondence expects is that for a given flat $GL_1$-bundle $\mathcal{E}$ on $X$, there exist a unique $\mathcal{D}$-module $\mathcal{F}_\mathcal{E}$ on $\text{Pic}(X)$ associated with the modification $h_x$. This correspondence is proven by P. Deligne. 

The general conjecture of the Langlands correspondence for a Lie group $G$ can be stated as follows.  We denote by $^LG$ the Langlands dual group of $G$. If $G=GL_1$, then its dual is isomorphic to $GL_1$. The set $\text{Loc}_{^LG}(X)$ of local systems is again identified with the set of conjugacy classes of representations $\rho:\pi_1(X)\to\hspace{-1mm}^LG$. And $\text{Pic}(X)$ is generalized to the moduli stack $\text{Bun}_G(X)$ of principle $G$-bundles on $X$. So the geometric Langlands correspondence implies that for a given flat $^LG$-bundle $\mathcal{E}$ on $X$, there is a unique $\mathcal{D}$-module $\mathcal{F}_{\mathcal{E}}$, called a Hecke eigensheaf, defined on $\text{Bun}_G(X)$ associated with the Hecke modification.

\subsection{Hecke Eigensheaf, Landau Level, and Anderson Localization} 
In this section we give a physical explanation about Hecke eigensheaves. The sections 4.3$\sim$4.5 in \cite{Frenkel:2005pa} will be helpful for more information. For this purpose, we physically interpret a sheaf. We are interested in a sheaf $\mathcal{S}=(\mathcal{S},\pi, B)$ whose fiber $\mathcal{S}_p=\pi^{-1}(p),~p\in B$ is a vector space. The dimension of fibers may differ at points. A standard example is the skyscraper sheaf $\mathcal{O}(x)$, which is a sheaf supported at a single point $x\in B$. How will they come into play in our story? First of all, sections of our sheaf are wave functions and the base space is the Riemann (complex energy) surface \cite{PhysRevB.48.11851,PhysRevLett.71.3697}. In comparison with the Landau levels, our sheaf can be visualizable as follows. 
\begin{figure}[H]
\centering
\includegraphics[width=10cm]{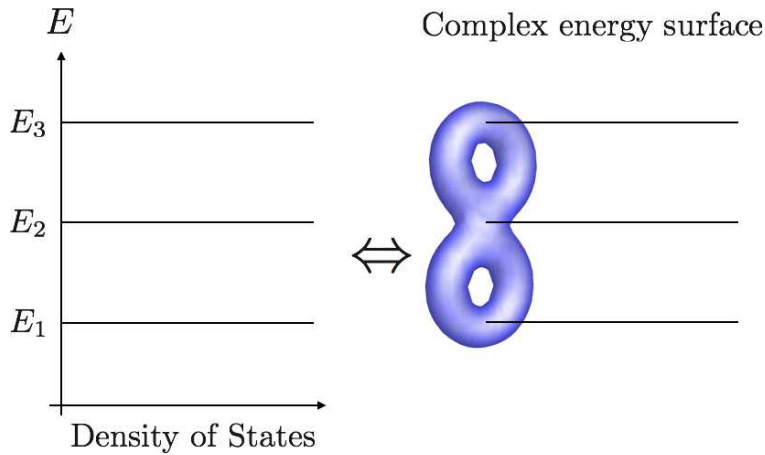}
\caption{\label{la}Landau level (left) and sheaf (right)}
\end{figure}
On the left side of the figure \ref{la}, the blank zones between the Landau levels show there are no eigenstates of the Hamiltonian, and the only states living in the Landau levels contribute to the quantum Hall effect. On the right side, the eigenstates forms bundles on the complex energy surface and the other states vanish. So the set of "complexified" Landau levels can be recognized as a sheaf $\bigcup_{i}\mathcal{O}(E_i)$.   

Now we are ready to explain Hecke eigensheavs. Note that each of the Landau levels in the figure \ref{la} is linear, therefore only one-dimensional momentum $k_y$ is a good quantum number. So one can define the only one-dimensional Berry connection $A_{k_y}$. Hence we may regard it as a flat connection\footnote{This flat connection, which is a Berry connection of $U(1)$-bundle, is a different one we used for a Wilson loop $W: \pi_1(T^2_{BZ})\to\hspace{-1mm}^LU(1)$.} on $T^2_{BZ}$, by setting $A_{k_x}=0$. A Hecke eigensheaf on $\text{Pic}^0(X)=\left\{\mathcal{L}\in\text{Pic}(X):0=\int_{X}\mathcal{L}\right\}$ is a $\mathcal{D}$-module of such flat connections. Of course this is not the whole story. Actual energy bands are "wavy" as shown with pictures in \cite{PhysRevLett.71.3697,PhysRevB.48.11851}, and the wavy parts possess nontrivial Chern numbers. This is the mechanism of Hecke modifications $\mathcal{L}\to\mathcal{L}(x)=\mathcal{L}\otimes\mathcal{O}(x),~x\in T^2_{BZ}$. 

Moreover the existence of plateaus in the figure \ref{hall} can be described by Hecke eigensheaves as shown in the figure below. If there exist impurity potentials in the system, wave functions localize around the potentials (in the real space). This phenomenon is called the Anderson localization \cite{PhysRev.109.1492}. As a result, each of the Landau levels becomes wide. However, the localized wave functions do not carry non trivial Chern numbers and only the extended wave function living in the original Landau level contributes to the Hall conductance \cite{PhysRevB.23.5632}. This is why the Hall conductance has plateaus. In the language of the geometric Langlands correspondence, this can be explained by saying that the Berry connections associated with those localized wave functions are flat, and hence they form a $\mathcal{D}$-module.  
\begin{figure}[H]
\centering
\includegraphics[width=6cm]{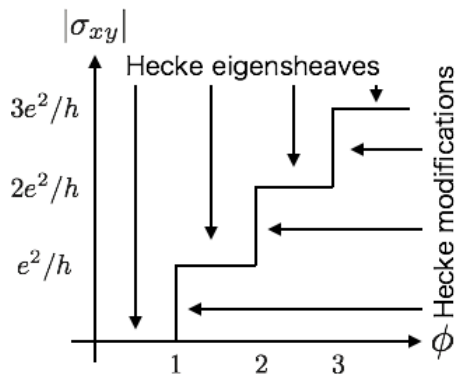}
\caption{\label{hall2}Hall conductance and Hecke eigensheaves}
\end{figure}

\subsection{Duality, Hecke operator, and K-theoretic view}
Algebraically, a vortex operator for $G$ is defined by a homomorphism $\varrho:U(1)\to G$, which is classified by highest weights of the dual group $^LG$ up to conjugation.  We write it in the most general way as $\varrho:e^{i\alpha}\to\text{diag}(e^{im_1\alpha},\cdots,e^{im_M\alpha})$, where $^Lw=(m_1,\cdots,m_M)$ is an $M$-plet of integers with $m_1\ge\cdots\ge m_M$, which is a highest weight of $^LG=U(M)$. As we have already seen, we obtain decomposition of the $U(M)$-bundle $\mathcal{E}$ into the sum of line bundles $\oplus_{i=1}^M\mathcal{L}_i$. Let $k_i$ be a singular point of $\mathcal{L}_i$. The vortex operator $V(^Lw)$ acts on $\mathcal{L}_j$ as $V(^Lw):\mathcal{L}_i\to \mathcal{L}_j\otimes \mathcal{O}(k_j)^{m_j}$, where $m_j$ is vorticity at $k_j$. In other words, it changes the Chern number $c(\mathcal{L}_j)$ by $m_j$. The total Hall conductance $\sigma_{xy}$ is the total Chern numbers of this system $\sigma_{xy}=\sum_{j=1}^M\sigma^j_{xy}$, which is the total vorticity of the system, in other words. 

By the way, the classification of vortex operators or Hecke operators is exactly the same as that of effective Hamiltonians. So far we have neglected contribution from conduction bands, and from now we suppose there are $N$ conduction bands and $M$ valence bands. So this system has $U(M+N)$ gauge group in general. Then effective Hamiltonians of the quantum Hall system is classified by the Grassmannian $Gr_{M, M+N}=U(M+N)/U(M)\times U(N)$ from the $K$-theoretic perspectives \cite{PhysRevB.78.195125}.  In terms of the geometric Langlands correspondence, the Hecke operators are classified as follows. Let $\mathcal{M},\mathcal{M}'\in \text{Bun}_{G}$ be principle $G=U(M+N)$ bundles on a Riemann surface $X$ such that $\mathcal{M}\subset \mathcal{M}'$ and $\mathcal{M}'/\mathcal{M}\simeq\mathcal{O}(x)^{M}$, where $x\in X$. As discussed, the Hecke operators modify the $G$-bundle $\mathcal{M}$ to $\mathcal{M}'$ at this singular point $x$, and it is known the space of such modifications is parametrized by points in $Gr_{M,M+N}$. (One may discover extra value in mathematical explanations \cite{Frenkel:2005pa} or in physical explanations \cite{Kapustin:2006pk,Witten:2015dta}).

\subsection{Chern-Simons Theory, $S$-duality and Mirror Symmetry}
Quantum Hall Effect has common description based on Chern-Simons theory, therefore it is meaningful to give some comments on the relation with the Langlands duality. We consider $2+1$-dimensional system which is parametrized by $x=(x^0,x^1,x^2)$, where $x^0$ stands for the time-direction and $x^1,x^2$ represent the space-directions. We may regard our system is product of $\mathbb{R}$ and a torus $T^2$ since our physics on the two dimensional space we have considered so far is periodic in the $x$ and $y$ directions respectively. 

Let $A$ be the background gauge filed of electromagnetism. The integer Hall conductance $\sigma_{xy}=\frac{k}{2\pi}$ is described by the Chern-Simons action 
\begin{equation}
S_{CS}=\int d^3x\frac{k}{4\pi}\epsilon^{\mu\nu\rho}A_\mu\partial_\nu A_\rho,
\end{equation}
whose $U(1)$-current is 
\begin{equation}
J^\mu=\frac{\partial S}{\partial A_\mu}=k\epsilon^{\mu\nu\rho}\partial_\nu A_\rho. 
\end{equation}
Especially this is nothing but the Hall current if one takes $\mu=x^1$, and the level $k$ corresponds to the bulk Hall conductance. 

The duality in quantum Hall effect that acts on the filling fraction $\nu$ can be understood as $S$-duality that acts on the inverse level $\hbar = 1/k$, which needs to be analytically continued away from integer values in order for the quality to be meaningful. When level $k$ is integer, the Langlands/$S$-duality formally maps $q = \exp (2 \pi i k)$ to $^Lq =\exp (2 \pi i / k)$. Continuing $q$ away from roots of unity is naturally accommodated in the complex Chern-Simons theory, which now indeed enjoys Langlands/$S$-duality \cite{Dimofte:2011jd}.
 
This duality describes the particle vortex duality \cite{Burgess:2000kj,PhysRevB.93.245151} and the geometric Langladns correspondence is easily understood. As we discussed in section \ref{GL}, the geometric Langlands correspondence states the duality between a Wilson loop and a Hecke operator. The Hecke operator corresponds to the vortex operator, which picks up the Chern number associated with the bulk Hall conductivity. To find the corresponding Wilson loop, we consider an anyon, which is a quasi-particle with magnetic flux.

The Landau level filling factor is defined by 
\begin{equation}
\nu=\frac{\text{The number of electrons in the system}}{{\text{The number of flux quanta passing through the system}}}.
\end{equation}
So $\nu$ is equivalent to the ratio of the number of electrons $n_e$ and flux quanta $\phi$ per placket. If $\nu=1/k$ there are $k$ flux quanta per electron. We may simply regard it as $\phi=k$ and $n_e=1$. The integer quantum Hall effect $\nu=k$ case can be regarded as $(\phi,n_e)=(1,k)$ or $(\phi,n_e)=(1/k, 1)$. If one prefers the former perspective, the duality $(\nu, \phi, n_e)\to(1/\nu,n_e, \phi)$ is similar to electric-magnetic duality as we see below soon. When one chooses the latter, the duality $\nu\to1/\nu$ along $\phi\to 1/\phi$ makes sense. This perspective is crucial for the duality in Hofstadter's butterfly as we discuss latter. 

We first investigate the case where $\nu$ is an integer (the Integer Hall effect) and will treat the fractional case latter. The modular group $SL(2,\mathbb{Z})$ acts on the complex Hall conductivity $\sigma=\sigma_{xx}+i\sigma_{xy}$ in such a way that
\begin{align}
S&:\sigma\to-\frac{1}{\sigma}\\
T&:\sigma\to\sigma+1
\end{align} 
So, with respect to $q=\exp(2\pi\sigma)$ and $^Lq=\exp(-2\pi/\sigma)$, the particle-vortex duality at $\sigma_{xx}=0$ in the integer quantum Hall system $(\sigma_{xy}=\nu=k)$ simply reads to $q\to\hspace{-1mm}^Lq$. Especially, on the plateau regions, where gauge connection is flat, the equation $\sigma_{xx}=0$ holds \cite{1978JPSJ...44.1839W,PhysRevLett.45.494} and the $S$-duality agrees with the Langlands duality of Wilson and Hecke operators. 

Let $\nu=k$, which means an unit flux $\phi_0$ is attached to an electron. The corresponding Wilson loop appears as an Aharonov-Bohm (AB) phase. Namely, the vector filed $\alpha$ which generates imaginary and negligibly thin magnetic flux attached to each electron should satisfy $\nabla\times\alpha=0$, however an electron moving around $\phi_0$ feels the vector potential and gains the phase $\exp(i2\pi\phi_0)$. From this picture, another way to understand a  Hecke operator, which is dual to the Wilson loop, is obtained as follows: the flux $\phi_0$ moving around $n_e=k$ electrons picking up the dual AB phase $\exp(i2\pi k\phi_0)$. This corresponds to the Hecke operator we discussed before as the vortex operator which pics up the Chern number associated with the Berry curvature. When we consider the vorticity, the location of the vortex depends on our gauge choice, however the former perspective of the AB-phase allows us to look the system in a gauge invariant way. By the way, this $S$-duality can be understood as so-called the particle-vortex duality and is consistent with the physical understanding of the geometric Langlands correspondence as electric-magnetic duality. By $S$-duality $\nu\to1/\nu$, the Wilson loop and the Hecke operators are exchanged and the Wilson loop has the phase $\exp(i2\pi k\phi_0)$. Therefore the geometric Langlands correspondence suggests the duality of the IQHE and the FQHE. 

The geometric Langlands correspondence is accompanied by the Lang- lands dual group $^LG$ of a given gauge group $G$. In our case $G=U(1)$ and hence $^LG=U(1)$.
The $\nu=1/k$ Laughlin state have an emergent $U(1)$ gauge field $a$, which is a global symmetry and dual to the background $U(1)$ gauge filed $A$. We begin with the Lagrangian 
\begin{equation}\label{aa}
\mathcal{L}[a]= -\frac{k}{4\pi}\epsilon^{\mu\nu\rho}a_\mu\partial_\nu a_\rho
\end{equation}
This Lagrangian can be generally written as 
\begin{equation}
\mathcal{L}_{CS}=\sum_{i,j}\frac{k_{i,j}}{4\pi}A_i\wedge dA_j,
\end{equation}
$k_{i,j}$ is known as the $K$-matrix. The $S$-duality is achieved by adding an off-diagonal part $\frac{1}{2\pi}\epsilon^{\mu\nu\rho}A_\mu\partial_\nu a_\rho$ to the Lagrangian \eqref{aa}
\begin{equation}
\mathcal{L}[a]\to \mathcal{L}'[A]=\mathcal{L}[a]+\frac{1}{2\pi}A\wedge da.
\end{equation}
Integrating out $a$ from $\mathcal{L}'[A]$, we obtain 
\begin{equation}
\mathcal{L}[A]=-\frac{1}{4\pi k}\epsilon^{\mu\nu\rho}A_\mu\partial_\nu A_\rho. 
\end{equation}
Immediately, we can see the Hall conductance $\sigma_{xy}$ is proportional to $1/k$. 

\if{
So to understand the IQHE-FQHE duality, consider the action  with Lagrange's multiplier $\gamma$ associated with Bianchi's identity $\epsilon^{\mu\nu\rho}\partial_\mu F_{\nu\rho}=0$
\begin{equation}
S=\int d^3x-\frac{1}{4}F_{\mu\nu}F^{\mu\nu}+\frac{1}{2}\gamma\epsilon^{\mu\nu\rho}\partial_\mu F_{\nu\rho},
\end{equation}
where $F_{\mu\nu}=\partial_\mu A_\nu-\partial_\nu A_\mu$. Suppose the boundary term banishes and integrate it with respect to $F$, we have $\partial^\mu\gamma=\frac{1}{2}\epsilon^{\mu\nu\rho}F_{\nu\rho}$. Substituting it into the original Lagrangian, we obtain 
\begin{equation}\label{chern}
S=\int d^3x\frac{1}{2}\epsilon^{\mu\nu\rho}\partial_{\mu}F_{\nu\rho}=2\pi\int_{S^2}\frac{1}{2\pi}F\in 2\pi\mathbb{Z},
\end{equation}
where $S^2$ is a two-sphere existing at the spacial infinity of $\mathbb{R}^{2,1}$. The form \eqref{chern} is nothing but the Chern number associated with the integer Hall conductance.

The Langlands/$S$-duality is summarized as 
\begin{align}
\begin{aligned}
S: G=U(1)&\to ^LG=U(1)\\
\mathcal{L}[a]&\to \mathcal{L}'[A]=\mathcal{L}[a]+\frac{1}{2\pi}A\wedge a\\
\nu=k&\to\nu=1/k
\end{aligned}
\end{align}
}\fi

Mirror symmetry and Chern-Simons theory is discussed in \cite{Dimofte:2011jd}, and here we leave a short summary which connects our perspective above. A central architecture in the study of Langlands/S-duality is Hitchin's moduli space $\mathcal{M}(G,C)$, which is the classical phase space of ($G$) Chern-Simons theory on a 3-manifold with a boundary Riemann surface $C$. If one consider the complexification of $G$ which we denote by $G_\mathbb{C}$ (e.g. $G=SU(2)$, $G_\mathbb{C}=SL(2,\mathbb{C})$), then $\mathcal{M}(G,C)$ is equivalent to the space $\mathcal{M}(G_{\mathbb{C}}, C)$ of flat $G_{\mathbb{C}}$ connections on $C$. The $S$-duality maps $\mathcal{M}_\text{flat}(G_\mathbb{C},C)$ to $\mathcal{M}_\text{flat}(^LG_\mathbb{C},C)$ and indeed they are a pair of mirror symmetry \cite{2003InMat.153..197H}. The particle-vortex duality can be described by the Landau-Ginzburg theory and, in supersymmetric situations, its analogue is referred to as mirror symmetry. Therefore, as long as we focus on the Langlands correspondence, the $S$-duality picture we have discussed in this work would be essentially same as the $S$(or $T$)-duality picture discussed in \cite{Kapustin:2006pk,Witten:2015dta} as mirror symmetry.

\section{Hofstadter's Butterfly and Langlands Duality of Quantum Groups}
Now we see that the Langlnads/$S$-duality also enhance our understanding on a different aspect of the quantum Hall effect. As we discussed, $\nu$ is equivalent to the ratio of the number of electrons $n_e$ and flux quanta $\phi$ per placket. If $\nu=1/k$ there are $k$ flux quanta per electron. We identify it with $\phi=k$ and $n_e=1$. The integer quantum Hall effect $\nu=k$ case can be regarded as $(\phi,n_e)=(1/k, 1)$. The duality $(\nu,\phi)\to (1/\nu,1/\phi)$ is crucial for the duality in Hofstadter's butterfly. On the $q$-parameter level, it states the duality between theories with $q$ and $^Lq$. As a well-known subject, the Hamiltonian \eqref{H} of the integer quantum Hall effect can be written by use of the quantum group $\mathcal{U}_q(sl_2)$ \cite{PhysRevLett.72.1890} and the Langlands duality of the quantum group endows with novel perspective on its fractal energy spectrum structure, known as Hofstadter's butterfly \cite{PhysRevB.14.2239}. 

\begin{figure}[h!]
\centering
\includegraphics[width=8cm]{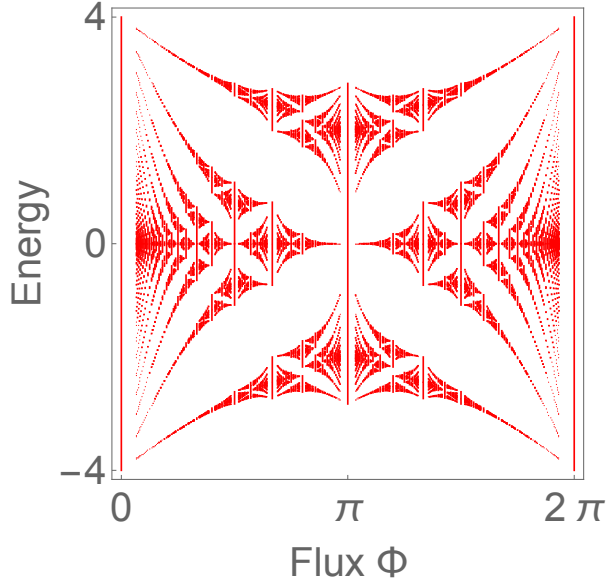}
\caption{Fractal energy spectrum, called Hofstadter's butterfly} 
\label{but}
\end{figure} 
It is known that this fractal spectrum is generated by the maps 
\begin{align}\label{a}
\begin{aligned}
(\phi,E)&\to (\phi+1, E)\\
(\phi,E)&\to (1/\phi, f(E)),
\end{aligned}
\end{align}
where $f$ is a some function. In \cite{Hatsuda:2016mdw}, this duality is described by a quantum geometric viewpoint and they relate the butterfly with the energy spectrum of relativistic Toda lattice. Tough the duality works for general $\phi$, we consider rational $\phi=P/Q$ to consider the butterfly ($P$ and $Q$ are co-prime). Now, our $q$-parameter is $q=\exp(i2\pi\phi)$. The duality is understood by the formula
\begin{equation}
P_{\phi}(E)=P_{1/\phi}(\widetilde{E}),~~\widetilde{E}=f(E). 
\end{equation}

A generic tight binding Hamiltonian we are interested in is 
\begin{equation}\label{tight}
H=\sum_{m,n}\left(c_{m+1,n}^\dagger c_{m,n}e^{iA^x_{m,n}}+R^2c_{m,n+1}^\dagger c_{m,n}e^{iA^y_{m,n}}+h.c.\right),
\end{equation} 
where $c_{m,n}~(c_{m,n}^\dagger)$ is the annihilation (creation) operator at $(m,n)$ site. When we choose the Landau gauge $A^x_{m,n}=0, A^y_{m,n}=2\pi m\phi$, it can be written as
\begin{equation}\label{t}
H=T_x+T_x^\dagger+R^2(T_y+T^\dagger_y),
\end{equation}
where we choose a $Q$-dimensional representation $\rho_Q$ of $\mathcal{U}_q(sl_2)=\{K^{\pm1},X^\pm\}$ with $q=e^{i2\pi P/Q}$ so that
\begin{align}\label{sl}
\begin{aligned}
T_x&=e^{ik_x}\rho_Q(X^+),~~T_y=e^{ik_y}\rho_Q(K)\\
\rho_Q(X^+)&=\begin{pmatrix}
0&1&0&\cdots&0\\
\vdots&\ddots&\ddots&\ddots\\
\vdots&&\ddots&\ddots&0\\
0&&&\ddots&1\\
1&0&\cdots&\cdots&0
\end{pmatrix}
,~~\rho_Q(K)=\text{diag}(q,q^2,\cdots,q^{Q})
\end{aligned}
\end{align}
These operators $T_x$ and $T_y$ are non commutative because of the Aharonov-Bohm phase for an electron moving around the flux:
\begin{equation}
T_xT_y=qT_yT_x. 
\end{equation} 
The energy spectrum consists of eigenvalues of this Hamiltonian, which is described by the Chambers relation \cite{PhysRev.140.A135}
\begin{equation}
\det(H(k,R)-E)=P_\phi(E,R)+h(k,R),~~k=(k_x,k_y)
\end{equation}
where $P_\phi(E,R)$ is a polynomial and $h(k,R)=2(-1)^{Q-1}(\cos(Qk_x)+R^{2Q}\cos(Qk_y))$. The energy spectrum displayed in Fig. \ref{but} satisfies the equation $P_\phi(E,R)=0$ under the mid band point condition $h(k_0,R)=0$, where $k_0=(\pi/2Q,\pi/2Q)$. Hence the anticipated formula $P_{P/Q}(E,R)=P_{Q/P}(\widetilde{E},\widetilde{R})$ implies the equivalence of the $Q$-dimensional representation \eqref{sl} of $\mathcal{U}_q(sl_2)$ and the $P$-dimensional representation of $\mathcal{U}_{^Lq}(sl_2)$, where $^Lq=e^{i2\pi/\phi}$ and $\mathcal{U}_{^Lq}(sl_2)$ is the Langlands dual quantum group of $\mathcal{U}_q(sl_2)$ \cite{Faddeev:1999fe, Frenkel2011}. We write this duality map by 
\begin{equation}\label{z}
S:(\mathcal{U}_q(sl_2), H)\to (\mathcal{U}_{^Lq}(sl_2), \widetilde{H}),
\end{equation}
where the dual Hamiltonian $\widetilde{H}$ is given by the following $P\times P$ matrix of the form 
\begin{equation}
\widetilde{H}=\widetilde{T}_x+\widetilde{T}_x^\dagger+\widetilde{R}^2(\widetilde{T}_y+\widetilde{T}^\dagger_y),
\end{equation}
where $\widetilde{T}_x=e^{i\widetilde{k}_x}\rho_P(X)$ and $\widetilde{T}_y=e^{i\widetilde{k}_y}\rho_P(Y)$. Since we expect the correspondence of the characteristic polynomials $\det(H-E)=\det(\widetilde{H}-\widetilde{E})$, we find $\widetilde{R}=R^{1/\phi}$ by comparing order of $R$ and $\widetilde{R}$ in $h(k,R)$ and $\widetilde{h}(\widetilde{k},\widetilde{R})$. 

This is consistent with the Langlands duality of quantum groups explained by the interpolating quantum group $\mathcal{U}_{q,t}(sl_2)$ \cite{Frenkel2011}, which is parametrized by arbitrary nonzero complex values $q,t$ and generated by $X^\pm,K^{\pm1},\widetilde{K}^{\pm{1}}$ such that
\begin{align}
\begin{aligned}
KX^\pm&=q^{\pm2}X^\pm K,~~\widetilde{K}X^\pm=t^{\pm2 }X^\pm \widetilde{K},\\
[X^+,X^-]&=\frac{K\widetilde{K}-(K\widetilde{K})^{-1}}{qt-(qt)^{-1}}.
\end{aligned}
\end{align}
The interpolating property of $\mathcal{U}_{q,t}(sl_2)$ appears as 
\begin{equation}
\mathcal{U}_{q,1}(sl_2)/\{\widetilde{K}=1\}\simeq \mathcal{U}_q(sl_2),~~\mathcal{U}_{1,t}(sl_2)/\{K=1\}\simeq \mathcal{U}_t(sl_2). 
\end{equation}
By definition, $\mathcal{U}_{q,t}(sl_2)$ is equivalent to the usual quantum group $\mathcal{U}_\varrho(sl_2)$ with generators $X^{\pm}, K\widetilde{K}$ and the parameter $\varrho=qt$. Taking $q=e^{i2\pi P/Q}$ and $t=\hspace{-2mm}~^Lq=e^{i2\pi Q/P}$, we find $\varrho=q~^Lq=e^{i2\pi(P/Q+Q/P)}$ is symmetric under exchanging $P$ and $Q$. The Langlands duality of quantum groups states that any irreducible representation of $\mathcal{U}_{q}(sl_2)$ would be $t$-deformed uniquely to a representation of $\mathcal{U}_{q,t}(sl_2)$ in such a way that its specialization at $q=1$ gives a representation of $\mathcal{U}_{t}(sl_2)$. The easiest case is $P=1$ and $Q=2$. A two-dimensional representation of $\mathcal{U}_q(sl_2)$ is dual to a one-dimensional representation of $\mathcal{U}_{^Lq}(sl_2)$, which is equivalent to $P_{1/2}(E,R)=P_{2/1}(\widetilde{E},\widetilde{R})$. Generically, we observe that a $Q$-dimensional representation of $\mathcal{U}_{q}(sl_2)$ and a $P$-dimensional representation of $\mathcal{U}_{^Lq}(sl_2)$ are dual. This explain the formula $P_{P/Q}(E,R)=P_{Q/P}(\widetilde{E},\widetilde{R})$.  

\if{
To connect $\phi$ and $\nu$, we consider a two dimensional square system of size $L\times L$ and suppose there are $N_e$ electrons. For the case $\phi=1/Q$, there are $Q$ energy bands and each of them contains $L^2/Q$ stats since there are $L^2$ one-particle states in total. On the other hand, magnetic flux per placket is $1/Q$ and the total magnetic flux is $N_\phi=L^2/Q$, which corresponds to the degeneracy of each of the Landau levels. By definition $\nu=\frac{N_e}{N_\phi}=\frac{N_e}{L^2/Q}\propto \frac{1}{\phi}$. So flipping $\nu\to 1/\nu$ is essentially equivalent to mapping $1/\phi\to \phi$. This is gives a more direct connection to the geometric Langlands duality associated with the Chern-Simons theory and the Langlands duality of the quantum group. 
}\fi

\if{ 
\section{Fractional Quantum Hall Effect and Langlands Program}
\if{
\subsection{Frobenius Operator}
Let $q$ be prime and $z$ be an element of $\mathbb{C}$. The Frobenius operator is defined by 
\begin{equation}
\text{Frob}_q(z)=z^q.
\end{equation}

The Landau level filling factor $\nu=1/q$ means that $q$ magnetic flux is attached to an electron. And the Frobenius operator comes into the FQHE as the Laughlin's wave function 
\begin{equation}
\Psi_{\nu=1/q}=\prod_{i<j}(z_i-z_j)^q\exp\left(-\frac{1}{4l^2}\sum_{i=1}^{N_e}|z_i|^2\right),\end{equation}
where $z_i$ stand for the coordinates of electrons.  Namely, the factor $q$ of $\text{Frob}_q$ corresponds to $1/\nu$. 

The corresponding Wilson loop is given by the Aharonov-Bohm phase. 
}\fi

\subsection{Knot, Chern-Simons, and Topological Quantum Computation}
Knots are objects in 3-space, however at first sight the Jones polynomial seems to be irrelevant to the background space. To interpret the Jones polynomial from three-dimensional geometry was a problem proposed by Atiyah very long ago, and it was solved by Witten \cite{witten1989}. Here we give a short introduction to this matter.  We consider a principle bundle $(E, \pi, ^LG, M)$ on an oriented 3-manifold $M$ without boundary, where $^LG$ is the Langlands dual group of a compact simple Lie group $G$ (we may take it to be simply connected if necessary). The Chern-Simons action is given by
\begin{equation}
S_{CS}(A)=\frac{1}{4\pi}\int_M\text{Tr}\left(A\wedge A+\frac{2}{3}A\wedge A\wedge A\right), 
\end{equation}  
where $A$ is a connection one-form (or a gauge field) on $E$. The partition function is defined by performing the Feynman path integral on the space $\mathcal{U}$ of connections:
\begin{equation}
Z_k(M)=\frac{1}{\text{vol}}\int_{\mathcal{U}}DA\exp[ikS_{SC}(A)],
\end{equation}
where $k\in\mathbb{Z}$ is the level of the theory. A knot $K$ in $M$ enters our story if we consider a Wilson loop of $A$.  We pic up an irreducible representation $^LR$ of $^LG$ and then the Wilson loop is given by 
\begin{equation}
W(K,^LR)=\text{Tr}_{^LR}P\exp\left(-\int_KA\right). 
\end{equation}

For simplicity we consider the case $G=SU(2)$ and $^LR$ is its fundamental representation. Then the vacuum expectation value of the Wilson loop 
\begin{equation}
Z_k(M=S^3, K, ^LR)=\frac{1}{\text{vol}}\int_{\mathcal{U}}DA\exp[ikS_{SC}(A)]W(K,^LR)
\end{equation}
corresponds to the Jones polynomial evaluated at $q=\exp\left(\frac{2\pi i}{k+2}\right)$. 
}\fi

\section{Finale}
Now we conclude this article with some comments. We successfully understand the Langlands philosophy in terms of the integer quantum Hall system. Our discussions can be summarized in the following dictionary: 
\begin{align}
\begin{aligned}\notag
\fbox{\text{Langlands Program}}&\leftrightarrow\fbox{\text{Quantum Hall effect}}\\
\text{Geometric Langlands}&\leftrightarrow\text{IQHE/FQHE}\\
\text{Quantum groups' duality}&\leftrightarrow\text{Hofstadter's butterfly}
\end{aligned}
\end{align}

This article is the first penguin for exploring topological insulators from a viewpoint of the Langlands program, and doors for further adventures are always open. We naively expect that many similar phenomena observed or expected in generic topological insulators will be addressed in the same way as we discussed. Moreover the theory of Anderson's localization, which distinguishes metal-insulator transitions, is one of the essential and cross-cutting issues in topological insulators. The key ingredients are the topological terms (the Wess-Zumino-Witten (WZW) terms) associated with the non linear sigma models \cite{PhysRevB.78.195125}. And the geometric Langlands correspondence manifests power for investigating the WZW model \cite{Frenkel:2005pa}. This suggests that mathematical background of topological insulators would be much more fruitful than what it had been believed.  

It is interesting to build another connection to the work done by Kapustin and Witten \cite{Kapustin:2006pk}, where $\mathcal{N}=4$ super Yang-Mills theory is essential to explain the geometric Langlands correspondence via mirror symmetry and $S$-duality. We can seek for a likely scenario in string theoretical approaches to the quantum Hall effect (and topological insulators) \cite{Ryu:2010hc, Ryu:2010fe}, in which the two-dimensional quantum Hall effect is described by using the $D3$-brane, on which the $\mathcal{N}=4$ supper symmetry does live \cite{Witten:1995im}. Moreover a generic topological insulator can be explained by the corresponding $D$-brane configuration, hence it may attract a general interest to build more strong connections among the geometric Langlands, topological insulators, and the supper symmetric theory. 

Finally, 3-dimensional understanding on Jones polynomials, which describes knots, is given by the Chern-Simons theory \cite{witten1989}. Categorufication of the Jones polynomial is known as Khovanov homology \cite{1999math......8171K, 2001math......3190K, 2003math......2060K, 2004math.....11447K}. Its physical interpretation is also proposed by Gukov, Vafa, and Schwarz \cite{Gukov:2004hz} in the context of topological string theory.  More recently its connection with gauge theory has been considered  \cite{Mazzeo:2013zga, WItten:2011pz, Witten:2016qzs} and it is predicted that the Langlands correspondence based on the $\mathcal{N}=4$ super Yang-Mills theory plays a fundamental role to understand Khovanov homology. It is a challenging open problem to find a physically realistic explanation about the conjecture, since knot theory and Jones polynomial endow the FQHE with fundamental description of anyons.

\bibliographystyle{utphys}
\bibliography{Ref}
\end{document}